\documentclass[aps,prl,twocolumn,superscriptaddress,floats]{revtex4-1}
\usepackage{txfonts}
\usepackage{amssymb}
\usepackage{graphicx}

\begin{document}

\title{Neutron Scattering Measurements of Spatially Anisotropic Magnetic Exchange Interactions in Semiconducting K$_{0.85}$Fe$_{1.54}$Se$_{2}$ ($T_N$=$280$ K)
}
\author{Jun Zhao}
\email{zhaoj@fudan.edu.cn}
\affiliation{
State Key Laboratory of Surface Physics and Department of Physics, Fudan University, Shanghai 200433, China
}
\author{Yao Shen}
\affiliation{
State Key Laboratory of Surface Physics and Department of Physics, Fudan University, Shanghai 200433, China
}
\author{R. J. Birgeneau}
\affiliation{
Department of Physics, University of California, Berkeley, California 94720, USA
}
\affiliation{
Department of Materials Science and Engineering, University of California, Berkeley, California 94720, USA
}
\author{Miao Gao}
\affiliation{
Department of Physics, Renmin University of China, Beijing 100872, China
}
\author{Zhong-Yi Lu}
\affiliation{
Department of Physics, Renmin University of China, Beijing 100872, China
}
\author{D.-H. Lee}
\affiliation{
Department of Physics, University of California, Berkeley, California 94720, USA
}
\affiliation{
Materials Sciences Division, Lawrence Berkeley National Laboratory, Berkeley, California 94720, USA
}
\author{X. Z. Lu}
\affiliation{
State Key Laboratory of Surface Physics and Department of Physics, Fudan University, Shanghai 200433, China
}
\author{H. J. Xiang}
\affiliation{
State Key Laboratory of Surface Physics and Department of Physics, Fudan University, Shanghai 200433, China
}
\author{D. L. Abernathy}
\affiliation{
Neutron Scattering Science Division, Oak Ridge National Laboratory, Oak Ridge, Tennessee 37831-6393, USA
}
\author{Y. Zhao}
\affiliation{
NIST Center for Neutron Research, National Institute of Standards and Technology, Gaithersburg, Maryland, 20899, USA
}
\affiliation{
Department of Materials Science and Engineering, University of Maryland, College Park, Maryland, 20742, USA
}

\begin{abstract}
We use neutron scattering to study the spin excitations associated with the stripe antiferromagnetic (AFM) order in semiconducting K$_{0.85}$Fe$_{1.54}$Se$_2$ ($T_N$=$280$ K). We show that the spin wave spectra can be accurately described by an effective Heisenberg Hamiltonian with highly anisotropic in-plane couplings at $T$= $5$ K. At high temperature ($T$= $300$ K) above $T_N$, short range magnetic correlation with anisotropic correlation lengths are observed. Our results suggest that, despite the dramatic difference in the Fermi surface topology, the in-plane anisotropic magnetic couplings are a fundamental property of the iron based compounds; this implies that their antiferromagnetism may originate from local strong correlation effects rather than weak coupling Fermi surface nesting.
\end{abstract}


\maketitle


\begin{figure*}[t]
\includegraphics[width=16cm]{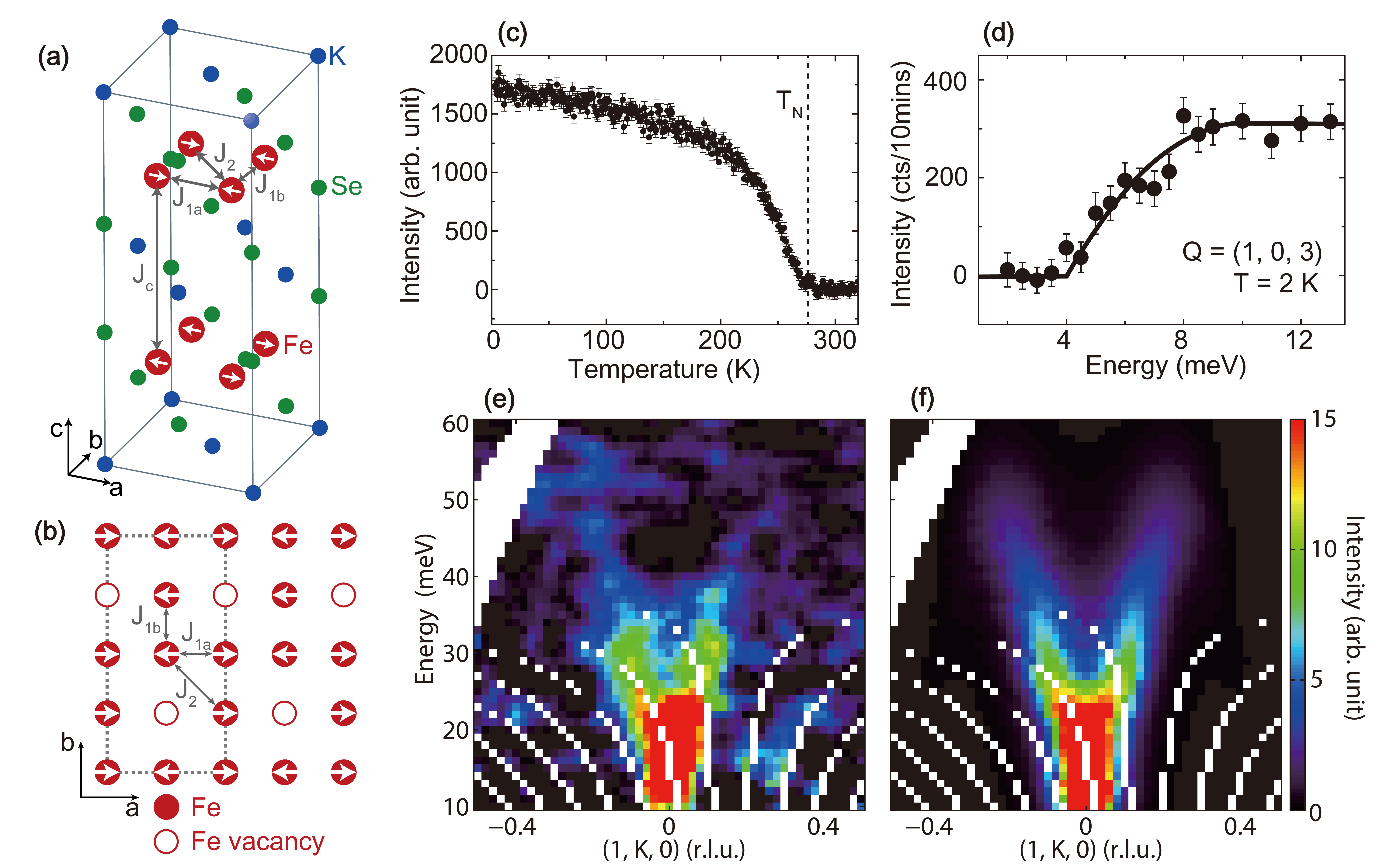}
\caption{(color online) Magnetic structure and spin excitations for semiconducting K$_{0.85}$Fe$_{1.54}$Se$_2$. (a), (b) Schematic diagram of the Fe spin ordering and Fe vacancies in K$_{0.85}$Fe$_{1.54}$Se$_2$. The dashed line indicates the magnetic unit cell. (c) Temperature dependence of the intensity of the magnetic Bragg peak associated with stripe AFM order. (d) Energy dependence of the spin excitation at ${\bf Q_{AFM}}$=($1,0,3$). (e) Background-subtracted magnetic excitations in K$_{0.85}$Fe$_{1.54}$Se$_2$ near ${\bf Q_{AFM}}$=($1,0$), measured on the ARCS time-of-flight chopper spectrometer with an incident energy of $E_i$ = $80$ meV. The white regions in the color plot are gaps between neutron detectors. (f) Calculated spin wave excitations using the model specified in the text. The error bars indicate one standard deviation throughout the Letter.
}
\label{cha}
\end{figure*}

The proximity of high-temperature superconductivity to antiferromagnetic (AFM) order in both cuprates and iron based materials has generated great interest in understanding the nature of their magnetism\cite{lee,johnston,dai,dagotto}. In close analogy to the layered copper oxygen planes of the cuprates, the square planar sheets of iron coordinated tetrahedrally by pnictogens or chalcogens are a common structural feature of the iron based superconductors. All parent compounds of the cuprate superconductors are antiferromagnets
defined by the same Heisenberg Hamiltonian \cite{lee}. Accordingly, it is important to identify the common magnetic properties (if any) of the iron based compounds. In fact, all of the parent compounds of the iron arsenides (such as 122-type $X$Fe$_2$As$_2$, where $X$=Ca, Sr, Ba) exhibit a stripe AFM order consisting of columns of parallel spins along the orthorhombic $b$ direction, along with the antiparallel spins along the $a$ direction [Fig. 1(a)] \cite{johnston,feng4,zhou2,note}. The stripe AFM wave vector connects the hole pockets at the zone center ($\Gamma$) and the electron pockets at the zone edges ($M$), implying that Fermi surface scattering may be involved in the magnetism of this system \cite{johnston}. The particularly notable feature associated with the stripe AFM order is that its spin wave spectra can be described with an effective Heisenberg Hamiltonian with highly anisotropic nearest-neighbor exchange couplings along the $a$ and $b$ axes ($J_{1a}$$>$0, $J_{1b}$$<$0), although the difference between the lattice constants $a$ and $b$ is rather small ($\sim1\%$) \cite{zhao1,diallo,leland,ewings}. Various theoretical scenarios have been proposed to explain this anisotropy, including the weak coupling model (itinerant), the strong coupling model (local), and a model with a combination of both itinerant electrons and local moments \cite{han,kaneshita,wysocki,stanek,si1,rpsingh,nevidomskyy,hpark,lv,chen,singh,kruger,fang}. As yet, the microscopic origin of the anisotropic magnetic coupling associated with the stripe antiferromagnetic order remains an issue of controversy.

The newly discovered alkali metal-intercalated iron selenide superconductors $A$$_x$Fe$_{2-y}$Se$_2$ ($A$=K, Rb, Cs), which have a crystal structure similar to that of 122 iron arsenides [Fig. 1(a)], provide a new testing ground for understanding the magnetism of the iron based materials. Intriguingly, neutron diffraction data also revealed a stripe AFM order in semiconducting K$_{0.85}$Fe$_{1.54}$Se$_2$ with potassium and iron vacancies. In this material, in contrast to the iron arsenides where the hole and electron pockets are reasonably well nested, the top of the hole band ($\Gamma$) is a few dozens of meV lower than the bottom of the electron band ($M$). Moreover, superconducting $A$$_x$Fe$_{2-y}$Se$_2$ only has electron bands crossing the Fermi energy ($E_F$) while the hole bands completely sink below $E_F$ \cite{xiang,gao,zhao2,feng1,feng2,zhou,ding}; the superconductor could be viewed as an electron-doped version of semiconducting K$_{0.85}$Fe$_{1.54}$Se$_2$. The different electronic structures between stripe AFM-ordered K$_{0.85}$Fe$_{1.54}$Se$_2$ and $X$Fe$_2$As$_2$ raise an important question as to whether the fundamental magnetic interactions that drive the stripe AFM order in these two systems are also different. This information may provide an important benchmark for the aforementioned theories describing the stripe antiferromagnetism of iron based materials \cite{han,kaneshita,wysocki,stanek,si1,rpsingh,nevidomskyy,hpark,lv,chen,singh,kruger,fang}.

In this Letter, we report inelastic neutron scattering studies of spin excitations in the stripe AFM-ordered semiconducting K$_{0.85}$Fe$_{1.54}$Se$_2$ ($T_N$= $280$ K; moment $\sim$ $2.8 $ $\mu_B$). The observed sharp and steeply dispersive spin waves can be described accurately by an effective Heisenberg Hamiltonian with anisotropic in-plane exchange couplings ($SJ_{1a}$=$37.9\pm 7.3$, $SJ_{1b}$=$-11.2\pm4.8$, $SJ_2$= $19.0\pm 2.4$, $SJ_c$= $0.29\pm 0.06$ meV) at low temperature ($T$$\ll$$T_N$). At high temperatures above $T_N$, the spin waves are replaced by quasi-two-dimensional short-ranged magnetic correlation with anisotropic dynamic spin correlation lengths in the $ab$ plane. These results suggest that the magnetisms in both semiconducting K$_{0.85}$Fe$_{1.54}$Se$_2$ and semimetallic $X$Fe$_2$As$_2$ have similar characteristics, and that their stripe AFM order may arise from superexchange
interactions between local moments driven by strong electron correlations.

Previous experiments have shown that single crystals of $A$$_x$Fe$_{2-y}$Se$_2$ have vacancies on iron or alkali-metal sites and tend to be phase separated, consisting of a $\sqrt{5} \times \sqrt{5}$ iron vacancy ordered block AFM insulating phase and superconducting or semiconducting phases with no $\sqrt{5} \times \sqrt{5}$ AFM order \cite{zhao2,feng1,li}. For the current experiment, one large semiconducting K$_x$Fe$_{2-y}$Se$_2$ single crystal ($8.5$ g) was grown with the Bridgman technique. Our neutron diffraction refinements on a small piece cleaved from the same crystal show that the fractions of the $\sqrt{5} \times \sqrt{5}$ block AFM-ordered insulating phase (K$_{0.8}$Fe$_{1.6}$Se$_2$) and the stripe AFM-ordered semiconducting phase (K$_{0.85}$Fe$_{1.54}$Se$_2$) amount to $\sim$ $75\%$ and $\sim$ $25\%$, respectively \cite{zhao2}. The stripe AFM-ordered K$_{0.85}$Fe$_{1.54}$Se$_2$ has rhombus iron vacancy order in the background of the $122$ structure, as illustrated in Figs. 1(a) and 1(b). Accordingly, we define the wave vector $Q$ at ($q_x$, $q_y$, $q_z$) as ($h$, $k$, $l$)= ($q_xa/2\pi$, $q_yb/2\pi$, $q_zc/2\pi$) reciprocal lattice units (r.l.u.) in the orthorhombic unit cell to facilitate the comparison with the 122 iron arsenides, where $a$=$b$=$5.527(3)$, and $c$= $14.07(6)$ $\AA$ are the lattice parameters at $5$ K. The temperature dependence of the magnetic Bragg peak in our large crystal indicates that the N$\rm \acute{e}$el temperature of the stripe AFM order is about $280$ K [Fig. 1(c)], consistent with the data obtained from the smaller crystals in Ref. \cite{zhao2}.

\begin{figure*}[t]
\includegraphics[width=16cm]{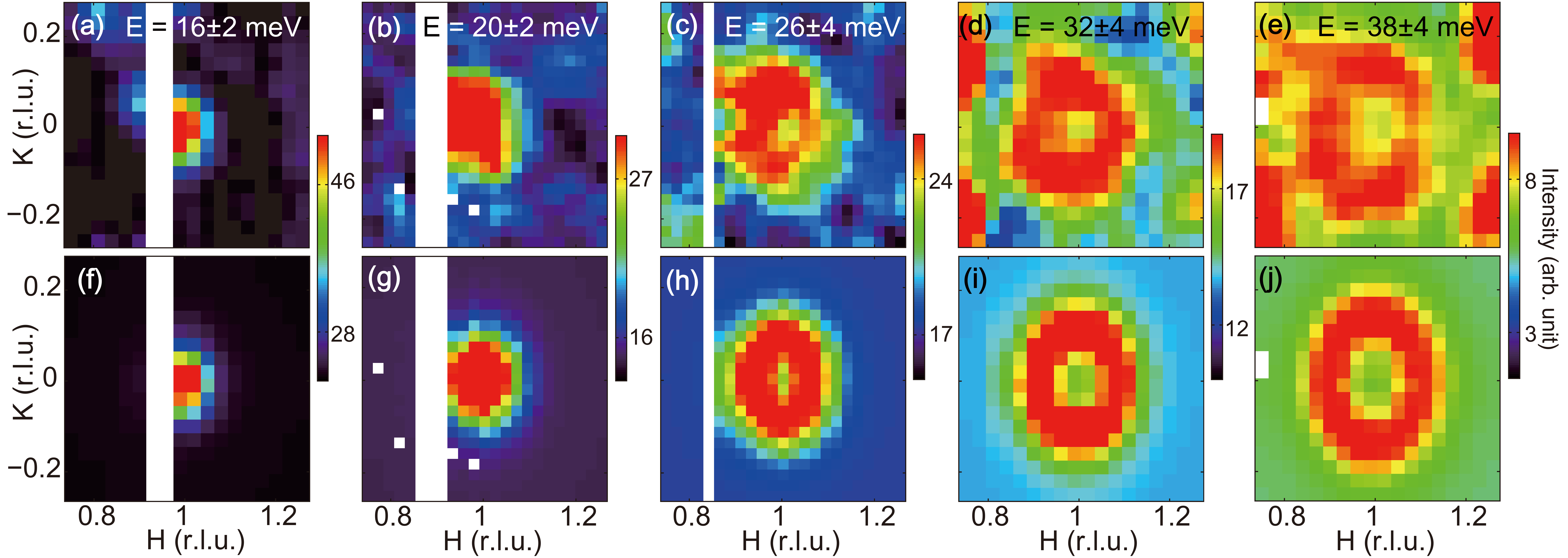}
\caption{(color online) (a)-(e) Constant energy slices through the spin waves in K$_{0.85}$Fe$_{1.54}$Se$_2$ at $5$ K, as observed on ARCS near $\bf{Q_{AFM}}$=($1, 0, L$),(a)  $L=1.75$; (b) $L=2.15$; (c) $L=2.8$; (d) $L=3.5$; (e) $L=4.2$. The spectrum is almost independent of $L$ above $16$ meV because of the weak coupling along the $c$ axis. (f)-(j) Resolution-convoluted simulation of the anisotropic Heisenberg model using the best fit parameters specified in the text. Each simulated slice is on the same intensity color scale as the measured slice at the same energy.
 }
\label{cha}
\end{figure*}

Our inelastic neutron scattering experiments were carried out on the ARCS time-of-flight chopper spectrometer at the Spallation Neutron Source of Oak Ridge National Laboratory and the BT-7 thermal triple axis spectrometer at the NIST Center for Neutron Research. For the ARCS experiment, the single crystal is aligned with the incident neutron beam parallel to the $c$ axis. For the
BT-7 measurements, we fixed the final neutron energy at $E_f$=$14.7$ meV and used pyrolytic graphite (PG) (0,0,2) as monochromator and analyzer.

Figure 2 illustrates a series of typical constant energy slices near the stripe AFM wave vector $\bf{Q_{AFM}}$=($1, 0, L$). At low energies ($< 20$ meV), the magnetic excitations are strongest at $\bf{Q_{AFM}}$. With increasing energy, the magnetic excitations disperse out of $\bf{Q_{AFM}}$. The ringlike scattering pattern is clearly anisotropic and elongated along the $K$ direction at high energies; this is reminiscent of the spin response of stripe-ordered $X$Fe$_2$As$_2$. In order to reveal the dispersion of magnetic excitations in momentum and energy space in more detail, we make a projection of the spin excitations along the $K$ direction near ($1$, $0$) with the background subtracted. The outcome immediately reveals a cone-shaped scattering arising from ($1, 0$) with energy extending up to $60$ meV [Fig. 1(e)]. Remarkably, both the spin excitation intensity and dispersion can be described accurately by a Heisenberg Hamiltonian with highly anisotropic exchange couplings [Fig. 1(f)] (discussed below).

To quantitatively determine the spin excitation dispersion relation in the $ab$ plane, we display, in Figs. 3(a)-3(e), cuts through the two-dimensional slices in Fig. 2 at different energies. A single peak centered on $\bf {Q_{AFM}}$ = ($1, 0$) at low energies evolves into a pair of peaks at $\sim 30$ meV, and continues to disperse outward at higher energies. Unlike the behavior in $X$Fe$_2$As$_2$ where the spin-wave damping increases appreciably with increasing energy \cite{zhao1,diallo,leland,ewings}, the magnetic excitations in K$_{0.85}$Fe$_{1.54}$Se$_2$ are sharp and essentially resolution limited at the energies probed. This behavior is consistent with the semiconducting nature of the ground state where spin-wave damping caused by itinerant electrons should be negligible.

The dispersion of the spin waves along the $c$ axis is determined by constant energy scans along the $L$ direction measured on the BT-7 triple-axis spectrometer [Figs. 3(g)-3(j)]. The spin excitation is observed to develop into a pair of peaks at very low energies ($\sim$ $10$ meV), indicating rather weak $c$ axis coupling. A constant energy scan along the $H$ direction near $\bf {Q_{AFM}}$ at $3$ meV is featureless, which suggests the presence of a single ion anisotropy gap [Fig. 3(f)]. A constant $Q$ scan at ($1,0,3$) further reveals that the magnitude of the gap is about $8$ meV [Fig. 1(d)], similar to the spin gaps $X$Fe$_2$As$_2$ \cite{zhao3,mcqueeney,matan}.

To describe the wave vector and energy ($Q$, $E$) dependence of the spin wave intensities, we adopt the spin Heisenberg model with the nearest-($J_{1a}$, $J_{1b}$, $J_c$) and next-nearest-neighbor ($J_2$) exchange couplings between the Fe moments,
\begin{equation}\label{eq:Ham}
\hat{H}=J_{1a}\sum_{\langle i,j_a \rangle}\vec{S}_i\cdot\vec{S}_{j_a} +J_{1b}\sum_{\langle i,j_b \rangle}\vec{S}_i\cdot\vec{S}_{j_b} +J_{c}\sum_{\langle i,j_c \rangle}\vec{S}_i\cdot\vec{S}_{j_c} + J_2\sum_{ \ll ij \gg}\vec{S}_i\cdot\vec{S}_j,
\end{equation}
where $\langle i,j_a \rangle$, $\langle i,j_b \rangle$, $\langle i,j_c \rangle$ and $\ll ij \gg$ signify the summations over the nearest neighbors along the $a$, $b$, $c$ axis and the next-nearest neighbors in the $ab$ plane, respectively. Similar to $X$Fe$_2$As$_2$, the crystal has two equally populated orthogonal twin domains in the $ab$ plane and only one twin domain is being probed near ($1$,$0$,$L$); this has been accounted for in our model. The spin wave excitations associated with the stripe AFM order with rhombus iron vacancies have been solved analytically using the equation of motion method in the framework of linearized spin wave theory \cite{gao}. We fitted the measured intensity of the spin-wave excitations and their dispersions by convoluting the model in Ref. \cite{gao} with the instrument resolution, using the TOBYFIT program \cite{toby}. Because of the presence of the rhombus iron vacancy order, there are six iron ions and two iron vacancies in one magnetic unit cell [Fig. 1(b)] and therefore, there should be three doubly degenerate branches of the spin waves associated with the stripe AFM order: one steeply dispersive acoustic mode (the gapless Goldstone mode) and two less-dispersive gapped optical modes. The optical modes are very weak and can not be clearly distinguished from the background in the current measurements. By fitting the most intense and dispersive acoustic spin wave in the whole Brillouin zone with the Heisenberg model, we were able to completely determine the effective exchange coupling constants: $SJ_{1a}$=$37.9\pm 7.3$, $SJ_{1b}$=$-11.2\pm4.8$, $SJ_2$= $19.0\pm 2.4$, $SJ_c$= $0.29\pm 0.06$ meV. The highly anisotropic exchange coupling constants ($J_{1a}>0>J_{1b}$) are very close to those reported in $X$Fe$_2$As$_2$ (Table 1). We note that the acoustic spin wave velocity in K$_{0.85}$Fe$_{1.54}$Se$_2$ is presumably lower than the velocities of $X$Fe$_2$As$_2$ because of the presence of iron vacancies. We also used Monte Carlo simulations to estimate the N$\rm \acute{e}$el temperatures based on the exchange coupling constants in Table 1. Interestingly, the calculated N$\rm \acute{e}$el temperatures are close to the measured temperatures and both show nonmonotonic variation with alkali or alkaline earth ion radius (Table I).

\begin{table}
\footnotesize
\caption{The magnetic exchange coupling constants and N$\rm \acute{e}$el temperatures of stripe AFM-ordered iron based compounds \cite{zhao1,leland,ewings}. The $T_N$ and $T'_N$ are measured and  Monte Carlo-calculated N$\rm \acute{e}$el temperatures, respectively. We note that the exchange coupling constants in SrFe$_2$As$_2$ at low energy (L) and high energy (H) are slightly different, which could be attributed to the involvement of the itinerant electrons in the magnetism \cite{ewings}. \label{tab:table1}%
}
\begin{ruledtabular}
\begin{tabular}{ccccccc}
\textrm{Compounds}&
\textrm{$SJ_{1a}$(meV)}&
\textrm{$SJ_{1b}$(meV)}&
\textrm{$SJ_{2}$(meV)}&
\textrm{$SJ_c$(meV)}&
\textrm{$T_N$/$T'_N$(K)}\\
\colrule
CaFe$_2$As$_2$ & $49.9\pm 9.9$ & $-5.7\pm4.5$ & $18.9\pm 3.4$ & $5.3\pm 1.3$ & 170/157\\
BaFe$_2$As$_2$ & $59.2\pm 2.0$ & $-9.2\pm1.2$ & $13.6\pm 1.0$ & $1.8\pm 0.3$ & 138/110\\
SrFe$_2$As$_2$(L) & $30.8\pm 1$ & $-5\pm4.5$ & $21.7\pm 0.4$ & $2.3\pm 0.1$ & 192/182\\
SrFe$_2$As$_2$(H) & $38.7\pm 2 $ & $-5\pm5$ & $27.3\pm 0.7$ & $2.3\pm 0.1$ & 192/212\\
K$_{0.85}$Fe$_{1.54}$Se$_2$ & $37.9\pm 7.3$ & $-11.2\pm4.8$ & $19.0\pm 2.4$ & $0.29\pm 0.06$ & 280/232\\
\end{tabular}
\end{ruledtabular}
\end{table}

\begin{figure}[t]
\includegraphics[scale=.14]{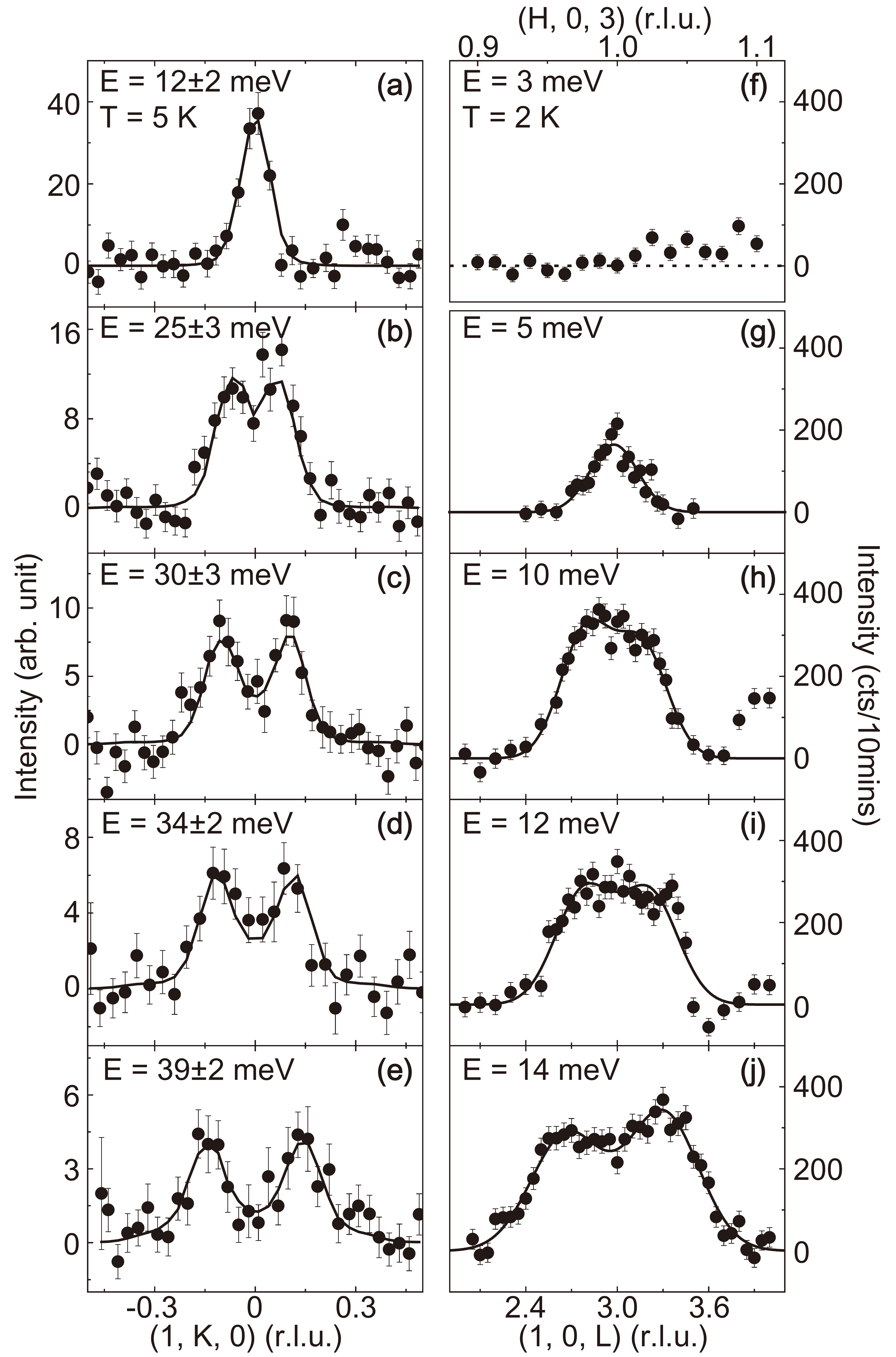}
\caption{(color online) Constant energy scans of the spin wave dispersion as a function of increasing energy at $5$ K fitted by the Heisenberg Hamiltonian described in the main text. (a)-(e) Constant energy cuts near ($1$,$0$,$L$) through the slice in Fig. 2 along the $K$ direction. (f)-(j) Constant energy scans along the $H$ or $L$ direction measured on the BT-7 triple axis spectrometer.
}
\end{figure}

\begin{figure}[t]
\includegraphics[scale=.14]{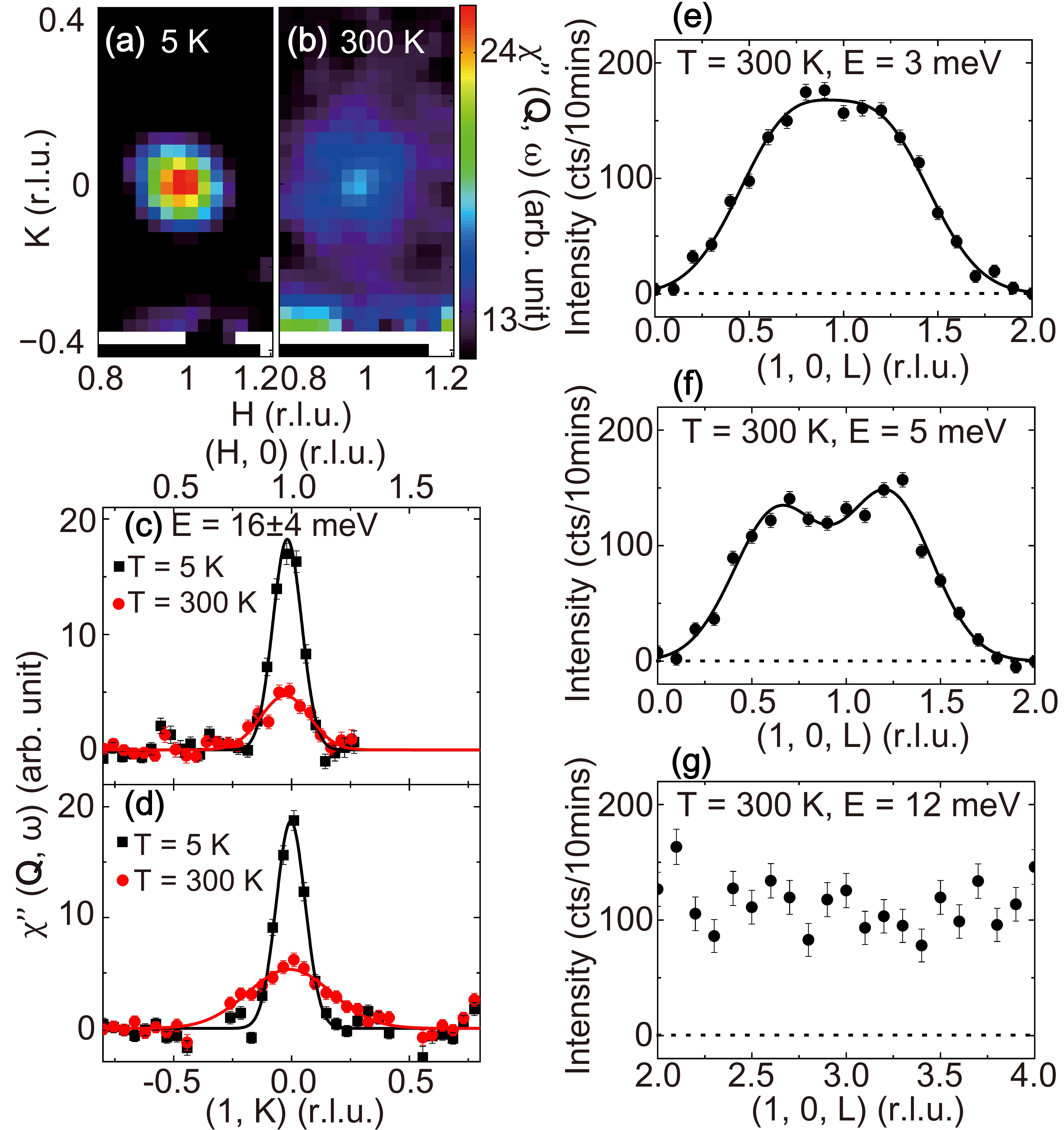}
\caption{(color online) Comparison of magnetic excitations in K$_{0.85}$Fe$_{1.54}$Se$_2$ at $5$ K and $300$ K.  (a)-(b) Constant energy ($E=16$ meV) slices at ($1$, $0$, $L$) at both $5$ K and $300$ K. (c)-(d) Constant energy cuts through Figs. 4(a) and 4(b) along the $H$ and $K$ directions. (e)-(g) Constant energy scans measured on the BT-7 spectrometer at ($1$, $0$, $L$) along the $L$ direction at $300$ K. The dashed lines indicate the background.
}
\end{figure}

In order to glean more insight into the nature of the stripe magnetic correlations, we studied the temperature dependence of the spin response in the semiconducting K$_{0.85}$Fe$_{1.54}$Se$_2$. Unlike the well-defined spin waves observed at $T\ll$$T_N$, the spin excitation becomes much broader at $T$=$300$ K [Figs. 4(a),4(b)]. The constant energy cuts through the $H$ and $K$ directions further reveal that the spin excitation is anisotropic in the $ab$ plane with dynamical spin correlation length in the $H$ direction much longer than that of the $K$ direction; this is consistent with the underlying stripe AFM structure Figs. 4(c),4(d)]. To determine the dynamical spin correlations out of plane, we performed constant energy scans along the $L$ direction [Figs. 4(e)-4(g)]. The clear peak feature revealed by the $L$scan at $3$ meV immediately suggests that the spin gap is absent at $300$ K. Moreover, the magnetic excitation develops into a pair of peaks at higher energy ($E=5$ meV), and eventually evolves into two-dimensional rodlike scattering at $12$ meV, suggesting that the effective exchange coupling along the $c$ axis is further reduced on warming to above $T_N$. These results are very much like the behavior of quasi-two-dimensional paramagnetic excitations in the $122$ iron arsenide parent compound $X$Fe$_2$As$_2$ \cite{diallo2,leland,ewings}.

The striking similarity of the spin dynamics in semiconducting K$_{0.85}$Fe$_{1.54}$Se$_2$ and semimetallic $X$Fe$_2$As$_2$ is very intriguing, given the different electronic ground states. Apparently, the naive weak coupling Fermi surface nesting picture cannot readily explain the anisotropic magnetic couplings and the resulting stripe AFM order observed in semiconducting K$_{0.85}$Fe$_{1.54}$Se$_2$. In the framework of the strong coupling approach, it has been suggested that the stripe AFM correlations may arise from the exchange interactions between localized moments \cite{xiang,gao}. This is supported by our observation that the spin excitation spectra are well defined and can be described accurately by a Heisenberg model. However, first principle calculations suggest that in-plane magnetic exchange couplings $J_{1a}$ and $J_{1b}$ are both antiferroamgnetic, with a small anisotropy induced by iron vacancies \cite{xiang,gao}. This prediction is inconsistent with the sign-changing anisotropy that we observe. Furthermore, no orthorhombic lattice distortion was observed down to $5$ K in K$_{0.85}$Fe$_{1.54}$Se$_2$ within our instrumental resolution. Therefore, the highly anisotropic magnetic coupling could be related to other degrees of freedom, such as nematic ordering, biquadratic interactions, and orbital ordering between $d_{xz}$ and $d_{yz}$ orbitals \cite{wysocki,stanek,si1,lv,chen,singh,kruger,fang}. Indeed, recent x-ray absorption spectroscopy and ARPES measurements have discovered evidence of orbital ordering in BaFe$_2$As$_2$ and CaFe$_2$As$_2$, respectively \cite{kim,wang}. Theoretical calculations also suggest that such orbital ordering can give rise to sign-changing anisotropic magnetic couplings \cite{lv,chen,singh,kruger}. These observations indicate that the anisotropy of magnetic couplings in K$_{0.85}$Fe$_{1.54}$Se$_2$ is very likely due to orbital ordering.

To summarize, we have shown that the spin wave spectrum in semiconducting K$_{0.85}$Fe$_{1.54}$Se$_2$ can be described accurately by a Heisenberg model with highly anisotropic in-plane exchange couplings, which is in closely analogous to the exchange couplings of the semimetallic iron arsenide parent compounds. These results suggest that such anisotropy is a fundamental property of stripe AFM-ordered iron based compounds, and does not necessarily only appear under Fermi surface nesting. The common characteristics of stripe AFM correlations provide a good starting point to delve into the magnetic phase diagram of iron based superconductors.

We thank D. X. Yao for helpful discussions. J.Z. and Y.S. acknowledge the start-up support from Fudan University and NSFC (Grant No.11374059). The research at UC Berkeley is supported by the Director, Office of Science, Office of Basic Energy Sciences, U.S. Department of Energy, under Contracts No. DE-AC02-05CH11231, No. DE-AC03-76SF008, and No. DE-AC02-05CH11231. M.G. and Z.Y.L. are supported by NSFC (Grants No. 11190024 and No. 91121008). The research at Oak Ridge National Laboratory's Spallation Neutron Source is sponsored by the Scientific User Facilities Division, Office of Basic Energy Sciences, U. S. Department of Energy.



\end{document}